\documentclass{PoS}

\title{Combination of the top-quark mass measurements from the Tevatron and from the LHC colliders}

\ShortTitle{Tevatron and LHC top mass combinations}

\author{\speaker{Fr\'ed\'eric DELIOT}\\
       {\rm On behalf of the ATLAS, CDF, CMS and D0 Collaborations}\\
       CEA-Saclay, Irfu/SPP, France \\
       E-mail: \email{frederic.deliot@cea.fr}}

\abstract{
I present here the results for the combination of the top-quark mass ($m_t$) measurements at the Tevatron
and at the LHC. These combinations are based on measurements using up to 5.8~\ifb\ from CDF and D0 yielding
$m_t = 173.18 \pm 0.56 ({\rm stat}) \pm 0.75 ({\rm syst})$~GeV,
and up to 4.7~\ifb\ from ATLAS and CMS leading to
$m_t = 173.3 \pm 0.5 ({\rm stat}) \pm 1.3 ({\rm syst})$~GeV.
}

\FullConference{36th International Conference on High Energy Physics,\\
		July 4-11, 2012\\
		Melbourne, Australia}

\newcommand{\ttbar}{\ensuremath{t\overline{t}}}
\newcommand{\ljets}{\mbox{$\ell$+jets}}
\newcommand{\dil}{\mbox{$\ell\ell$}}
\newcommand{\met}    {\mbox{$\not\!\!E_T$}}

\newcommand{\qqbar}{\ensuremath{q\bar{q}}}
\newcommand{\ifb}{\mbox{fb$^{-1}$}}

\begin{document}

\section{Introduction}
The top quark is a unique particle . Indeed it is the heaviest known elementary particle and it has a Yukawa coupling
to the Higgs boson close to unity, which may indicate that it plays a special role in the 
electroweak symmetry breaking mechanism. It is the only quark that decays before hadronizing, which 
gives the unique opportunity to study a bare quark. At hadron colliders, it is mainly produced in
pairs by the quantum chromodynamic interaction (QCD) via quark-antiquark annihilation ($\qqbar \to \ttbar$, 
dominant at the Tevatron) or gluon fusion ($gg \to \ttbar$, dominant at the LHC).
Within the Standard Model (SM), the top quark decays almost 100\% of the time into a $W$ boson 
and a $b$~quark. The \ttbar\ signatures are therefore classified according to the decays 
of the $W$ bosons into the lepton+jets channel (\ljets), the dilepton channel (\dil), the alljets
channel and the \met+jets channel.

\section{The top-quark mass}
The top-quark mass is a free parameter of the SM. It should be measured experimentally with a
high precision because it enters into the computation of quantum loop corrections for several
observables. In particular the $W$ boson mass ($M_W$) receives radiative corrections proportional 
to $m^2_t$ and to the logarithm of the Higgs boson mass ($m_H$). Therefore measuring $M_W$ and $m_t$
allows to indirectly constrain $m_H$ and consequently to test the consistency of the SM by 
comparing direct and indirect
determination of $m_H$. The latest indirect constraint on $m_H$ through a fit of the electroweak precision
data yields: $m_H = 94^{+29}_{-24}$~GeV or $m_H < 152$~GeV at 95\% confidence level~\cite{ewfit}.  

There are three main methods to measure $m_t$ directly. The simplest method is called the template
method. It relies on a given observable in data sensitive to $m_t$ 
(which is often the reconstructed $m_t$ from the top-quark decay products) 
which is compared to distributions (templates) produced with Monte Carlo 
(MC) simulation generated with different $m_t$ input values.
The matrix element method is based on the construction of a per-event probability 
computed with the leading-order \ttbar\ matrix element using the full event kinematic
informations. Finally the ideogram method uses an event likelihood computed
as a Gaussian resolution function with a Breit-Wigner to model the top-quark signal.
For channels that contain at least one $W$ boson that decays hadronically,
the jet energy scale (JES) can be calibrated by constraining the invariant mass of the two
light jets to the world-average value of $M_W$.
This allows to limit the impact of the JES uncertainty on the measured $m_t$.
In order to correct for approximations and for any potential biases, all methods are calibrated
using MC samples.

\section{Top-quark combination at the Tevatron}

The Tevatron experiments have published several measurements of $m_t$ over the last twenty years 
using Run~I (1992-1996) and Run~II (2001-2011) data sets in 6 top-quark decay channels and different
measurement techniques. 
Twelve measurements that use up to 5.8~\ifb\ of data are used for the combination.
The best independent measurements per channel in each experiment are chosen, 
eight from CDF and four from D0.
The inputs are 5 \ljets\ measurements (CDF and D0, Run~II and Run~I, and a CDF Run~II result 
based on the decay length of B hadrons); 2 alljets measurements (CDF Run~II and Run~I); 
4 \dil\ measurements (CDF and D0, Run~II and Run~I); and a \met+jets measurement (CDF Run~II).

The combination uses the Best Unbiased Linear Estimate 
(BLUE)~\cite{BLUE-method-2,BLUE-method-3} method
that calculates the combined mass value $m_t^{\rm comb}$ (an estimator of the true top quark mass) as 
a linear weighted sum of the input results $m_t^i$:
$m_t^{\rm comb}  = \sum_{i=1}^{12} w_i \; m_t^i$,
where the weights are computed as:
\begin{equation}
w_i   = \frac{\sum_{j=1}^{12}
             {\rm Covariance}^{-1} \! \left(m_t^i,m_t^j\right)}
             {\sum_{i=1}^{12} \sum_{j=1}^{12}
             {\rm Covariance}^{-1} \! \left(m_t^i,m_t^j\right)},
\end{equation}
and ${\rm Covariance}^{-1} \! \left(m_t^i,m_t^j\right)$
are the elements of the inverse of the covariance matrix (i.e. the error matrix)
of the input measurements.
The uncertainties on the 12 inputs are separated into 15~parts to properly estimate
this covariance matrix getting the correct pattern of correlation between channels, run periods 
and experiments. After several years of discussion, the CDF and D0 collaborations have agreed on 
a common list of systematics, on common evaluations and splitting of the systematic uncertainties
and on their correlations.
This common splitting is described in the following.

The systematic uncertainties from JES are the largest sources of systematic uncertainties
on the measured $m_t$ inputs. These are split into 7 different parts:
\begin{itemize}
\item Uncertainty from light-jet response (1) (also called rJES): this part is specific to CDF 
measurements and comes from CDF method of calibrating JES using single-pion response in data and 
in MC. It is implemented by tuning the simulation and is assumed to be 100\% correlated between
all CDF measurements;
\item Uncertainty from light-jet response (2) (also called dJES): this part corresponds to the 
absolute and relative uncertainty on JES calibration using $\gamma$+jets events 
in D0 and $\eta$-dependent calibration in CDF. It is 100\% correlated within the same experiment 
and the same run period;
\item Uncertainty from out-of-cone corrections (also called cJES): this is the uncertainty coming
from out-of-cone corrections to MC showers for CDF and D0 Run~I measurements. 
It is 100\% correlated between all measurements;
\item Uncertainty from offset (also called UN/MI): this part arises from the uncertainty on
JES coming from the uranium decay noise and pile-up from previous collisions. 
Due to the smaller integration time for D0 calorimeter 
electronics in Run~II, it is only a significant source of uncertainty in D0 Run~I measurements.
It is 100\% correlated within D0 Run~I measurements; 
\item Uncertainty from modeling of $b$-jets (also called bJES): this part corresponds to 
the difference between models of $b$-jet hadronization. 
It is 100\% correlated between all measurements;
\item Uncertainty from the response difference for $b$-, $q$- and $g$-jets (also called aJES): it arises
from MC/data difference in response between $b$-jets, light and gluon jets. 
It is assumed to be 100\% correlated within the same experiment and the same run period; 
\item Uncertainty from in-situ light-jet calibration (also called iJES): this part is relevant for
channels with at least one $W$ boson decaying hadronically that uses the light dijet invariant mass
to calibrate the JES. It is 100\% uncorrelated between all measurements and
is scaling with the statistical uncertainty of the amount of analyzed data. 
\end{itemize}

The other sources of systematic uncertainties that are not related with JES are also split into 7 parts:
\begin{itemize}
\item Uncertainty from jet modeling: this part arises from the uncertainty on jet identification
efficiency and jet smearing at D0. It is 100\% correlated between all D0 Run~II measurements;
\item Uncertainty from lepton modeling: it comes from the uncertainty on electron and muon momentum scale,
including also the uncertainty on the muon momentum smearing at D0. It is assumed to be 100\%
correlated within the same experiment and the same run period;
\item Uncertainty from signal modeling: this part contains the uncertainty coming from the limited
knowledge on the parton distribution functions and on the $q \bar{q}/gg$ fraction in the \ttbar\ production, 
from corrections due to higher-order QCD, from uncertainties on the initial and final-state radiation
modeling, from the 
hadronization model and from color reconnection. It is 100\% correlated between all measurements;
\item Uncertainty from the multiple interaction model: it arises from the uncertainty on the modeling
of pile-up in the MC and is 100\% correlated within the same experiment and the same run period;
\item Uncertainty from the background coming from theory: this part contains the uncertainty on
the NLO fraction of heavy flavor jets in the $W$+jets MC samples, the uncertainty from factorization
and renormalization scales in the $W$+jets simulation and from theory cross sections used to normalize
the MC samples. It is assumed to be 100\% correlated between all measurements in the same channel;
\item Uncertainty from background estimation based on data: this uncertainty comes from the MC/data difference
in some background distributions and from the signal/background fraction. 
It is 100\% correlated within the same experiment and the same run period in the same channel;
\item Uncertainty from the calibration method: it arises from the uncertainty on the calibration curve
and is uncorrelated between all measurements.
\end{itemize}

With this splitting of the systematic uncertainties and the above correlations, the combined value
for the Tevatron top-quark mass is: 
$ m_t = 173.18 \pm 0.56 ({\rm stat}) \pm 0.75 ({\rm syst})$~GeV corresponding to a total uncertainty
of 0.94~GeV and a total relative error of 0.54\%~\cite{Aaltonen:2012ra}. 
The $\chi^2$ for this combination is 8.3 for 11 degrees
of freedom which is equivalent to a 69\% probability for agreement among the 12 input measurements.
Figure~\ref{fig:tevcombi} shows this combined value
together with the input measurements.
In the combination, the input measurements that get the largest weights are CDF and D0 Run~II 
\ljets\ measurements (55.5\% and 26.7\%) and CDF Run~II measurement in the alljets channel (14\%).
The main uncertainties in the combined value are the statistical uncertainty (0.56~GeV), 
the systematic uncertainty from in-situ light-jet calibration (0.39~GeV)
and from signal modeling (0.51~GeV). The two first ones will scale down when all the Tevatron 
data will be analyzed. We can expect therefore that the final combination of the top-quark 
mass measurement at the Tevatron will have a total uncertainty around 0.7-0.8~GeV.

\begin{figure}[!htb]
\begin{center}
\includegraphics[width=95mm]{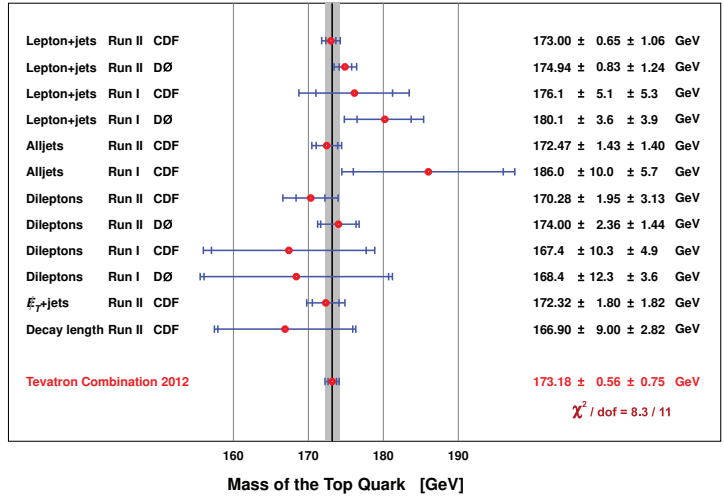}
\caption{The 12 input measurements of $m_t$ from the Tevatron collider experiments along with 
the resulting combined value~\cite{Aaltonen:2012ra}.}
\label{fig:tevcombi}
\end{center}
\end{figure}

\section{Top-quark combination at the LHC}

At the LHC, 7 input measurements are used in the combination.
All were obtained by analyzing LHC data at 7~TeV.
These inputs are ATLAS~\cite{atlas} 2010 measurement in the \ljets\ channel using 35~$pb^{-1}$,
ATLAS 2011 \ljets\ measurement using 1.0~\ifb, ATLAS 2011 measurement in the alljets channel
using 2.0~\ifb, CMS~\cite{cms} 2010 measurement in the \dil\ channel using 36~$pb^{-1}$, CMS 2010 \ljets\
measurement using 36~$pb^{-1}$, CMS 2011 \dil\ measurement using 2.3~\ifb\ and CMS 2011 measurement
in the $\mu$+jets channel using 4.7~\ifb.

The systematic uncertainties on these measurements are split using the same categories as 
described for the Tevatron in the previous section. However two categories are not used (aJES and
cJES, the latter being included into dJES) but two additional categories are taken into account. 
Since the uncertainties from
initial- and final-state radiations and from hadronization are assumed to be 50\% correlated between
the LHC experiments, these are separated from the signal modeling systematic uncertainty which is taken to be
100\% correlated. A new category also includes the uncertainty from the modeling of underlying event,
which was not considered as a separate source at the Tevatron but included in the uncertainty from 
signal modeling.

The LHC top-quark mass combination result yields: 
$ m_t = 173.3 \pm 0.5 ({\rm stat}) \pm 1.3 ({\rm syst})$~GeV corresponding to a total uncertainty
of 1.4~GeV~\cite{LHCcombi}.
The $\chi^2$ for this combination is 2.5 for 6 degrees
of freedom which is equivalent to a 86\% probability for agreement among the input measurements.
Figure~\ref{fig:lhccombi} shows this combined value
together with the input measurements at the LHC.
The largest weights in the combination are carried by the CMS 2011 $\mu$+jets measurement (65.7\%) 
and the ATLAS 2011 \ljets\ measurement (23.3\%).
The assumed correlations between the systematic uncertainties were checked by varying the fully correlated
sources from the default value of 100\% to 0\% simultaneously in steps of 10\%. This lead to a variation
of the combination central value of less than 200~MeV.
The largest uncertainties in the combination arise from the uncertainty on initial- and final-state 
radiation (0.69~GeV) and from color reconnection (0.55~GeV) as well as from uncertainty in the 
modeling and response from $b$-jets (bJES: 0.68~GeV).

\begin{figure}[!htb]
\begin{center}
\includegraphics[width=95mm]{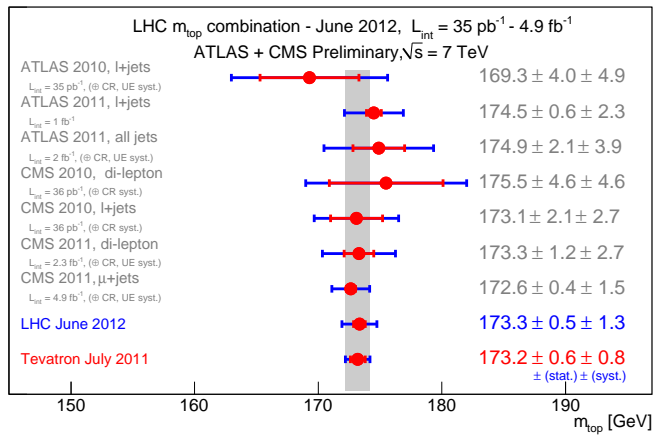}
\caption{Input measurements and result of the LHC combination~\cite{LHCcombi}.}
\label{fig:lhccombi}
\end{center}
\end{figure}

\section{Conclusion}
The combination of the top-quark mass measurements from the Tevatron and from the LHC were presented.
The Tevatron combination, which was first published in 2012, has now a total uncertainty below 1~GeV
while the uncertainty on the final combination is expected to be around 0.7 to 0.8~GeV.
The first preliminary combination of the LHC measurements has a total uncertainty of 1.4~GeV.
With the large amount of \ttbar\ statistics available at the LHC, this uncertainty is expected to 
decrease by constraining systematic uncertainties directly using data and
by performing top-quark mass measurements in specific regions of the phase space.

\end{document}